# POTENTIAL CUSTOMER LIFETIME VALUE IN FINANCIAL INSTITUTIONS: THE USAGE OF OPEN BANKING DATA TO IMPROVE CLV ESTIMATION

João B. G. de Brito[1][0000-0002-4645-5217], Rodrigo Heldt[1][0000-0002-3435-8962],
Cleo S. Silveira[1][0000-0001-5340-9107], Matthias Bogaert[2][0000-0002-4502-0764],
Guilherme B. Bucco[1][0000-0002-3813-8406], Fernando B. Luce[1][0000-0003-2589-2845],
João L. Becker[3][0000-0003-4176-7374], Filipe J. Zabala[1][0000-0002-5501-0877], and
Michel J. Anzanello[1][0000-0002-4421-7004]

[1] Federal University of Rio Grande do Sul, Porto Alegre, Brazil
[2] Ghent University, Ghent, Belgium
[3] Fundação Getúlio Vargas, São Paulo, Brazil

joaobatista.goncalvesdebrito@ugent.be

## ABSTRACT

Financial institutions increasingly adopt customer-centric strategies to enhance profitability and build long-term relationships. While Customer Lifetime Value (CLV) is a core metric, its calculations often rely solely on single-entity data, missing insights from customer activities across multiple firms. This study introduces the Potential Customer Lifetime Value (PCLV) framework, leveraging Open Banking (OB) data to estimate customer value comprehensively. We predict retention probability and estimate Potential Contribution Margins (PCM) from competitor data, enabling PCLV calculation. Results show that OB data can be used to estimate PCLV per competitor, indicating a potential upside of 21.06% over the Actual CLV. PCLV offers a strategic tool for managers to strengthen competitiveness by leveraging OB data and boost profitability by driving marketing efforts at the individual customer level to increase the Actual CLV.

## 1. INTRODUCTION

Historically, businesses prioritized brands and products over customer needs, limiting their ability to meet demands and build loyalty [1]. A shift to customer-centric strategies is essential, emphasizing CLV as a metric for fostering sustainable growth and aligning strategy with evolving customer expectations for profitability [2]. CLV measures a customer's value, aiding resource allocation, personalized marketing, and strategic decision-making [3]. This is particularly vital in sectors like banking, insurance, and subscriptions, helping identify high-value customers, optimize marketing, and deepen customer relationships for long-term success [4].

In finance, CLV drives personalized offerings, portfolio optimization, and improved loyalty amid competition [5]. Focusing on high-value customers using data-driven segmentation maximizes returns and sustains competitiveness [6]. However, traditional CLVs rely solely on single-entity data, missing insights from customer activities across multiple firms. OB transforms financial services by enabling data sharing with customer consent, offering a holistic view of interactions across institutions. Though OB adoption is still limited, our objective is to leverage Open Banking (OB) data to propose and empirically apply a framework to estimate Potential Customer Lifetime Value (PCLV), incorporating customers' transactions with competitors to provide a comprehensive assessment of Total CLV, which combines PCLV and Actual CLV [7]. By adopting the proposed framework, institutions gain insights into performing better in customer relationship management in an environment where competitiveness is expected to increase due to customer data sharing across competitors through the OB. It may be accomplished by using PCLV to drive marketing efforts such as add-on selling and retention strategies at the individual customer level to increase the Actual CLV.

## 2. THEORETICAL FOUNDATION

Corporate strategies have historically prioritized product-centric approaches, focusing on efficiency and innovation while often neglecting customer needs [2].

Although this approach has driven advancements, it has struggled to adapt to evolving customer expectations or create groundbreaking innovations that redefine markets [8]. These limitations have highlighted the need for a shift toward customer-centricity, emphasizing customer needs and experiences as central to business strategies and requiring changes in organizational culture, processes, and technologies [9].

CLV plays a vital role in this shift, quantifying the total value a customer contributes to a business over their relationship. By transitioning from transaction-based interactions to sustained, value-driven engagements, CLV enables organizations to align operations with customer priorities, enhancing satisfaction and maximizing long-term profitability [10]. As a strategic framework, CLV bridges customer-focused strategies with measurable outcomes, fostering deeper relationships and improved financial performance. CLV is particularly relevant in financial institutions because it relies on long-term relationships and diverse product portfolios. It helps identify high-value customers, tailor personalized services, and optimize resources, enhancing satisfaction and loyalty while supporting revenue forecasting and mitigating churn risks (Brito, Bucco et al., 2024; Fader et al., 2005). Incorporating CLV into planning drives customer retention, sustainable growth, and competitive advantage. CLV relies on future cash flows' net present value (NPV), accounting for retention probabilities and discount rates (Eq.1) [12].

$$Actual\ CLV_i = \frac{((cm_i * 12) * r_i)}{(1 + d - r_i)} \qquad (1)$$

Where $i$ is the customer; $cm_i$ is the contribution margin of the customer $i$ for the current month; $r_i$ is the probability of customer retention of customer $i$; and, $d$ is the annual discount rate.

However, this Actual CLV focuses solely on a company's interactions, overlooking potential value from customers' relationships with competitors. To address this gap, PCLV quantifies untapped business potential from competitor relationships [7], offering a more comprehensive view of customer value and identifying growth opportunities.

Operationalizing PCLV requires external data on customer transactions with competitors, a challenge that OB addresses. Initiated with the European Union's PSD2 in 2015, OB enables secure, standardized data sharing between financial institutions with customer consent [13]. Globally adopted, OB provides access to transaction, credit, and socio-demographic data, offering deeper insights into customer behavior [14].

OB uses APIs regulated by security standards to facilitate seamless data exchange, empowering institutions to tailor strategies while enhancing customer experiences [15]. Institutions that fail to adapt risk losing customers to competitors, leveraging OB data for personalized services [16]. By integrating CLV and OB, financial institutions can operationalize PCLV, boost retention, and maximize profitability, strengthening their competitive edge in a data-driven market.

## 3. METHOD

### 3.1 PCM

Estimating PCM is challenging due to variations in product and service organization across financial institutions. While competitors' offerings must align with the focal institution, Open Banking (OB) uses a standardized Product Service Category (PSC) interface for interoperability. PSC categorization, covering items like credit cards, auto loans, and mortgages, requires precise mapping to ensure prediction accuracy.

PCM prediction occurs in two phases. First, an XGBoost Regression model is trained to predict the total contribution margin per customer for the focal institution's PSCs. Second, using OB data on competitor transactions, the model predicts PCM for competitors' PSCs, simulating their migration to the focal institution. The standardized PSC format ensures OB data integrates seamlessly as input for the trained model.

### 3.1.1 Phase 1 – Model Training and Testing

Once the dependent and independent variables were defined, the predictive model was trained to accurately capture the patterns in the focal institution's data. The model was trained using the Extreme Gradient Boosting (XGBoost) Regression to predict the total contribution margin from all PSCs of the focal institution, according to Eq. 2.

$$\text{Total Contribution Margin Focal Institution}_i \sim \text{Amount of Credit Focal Institution}_{i,c} + \text{Amount of Credit National Financial System}_{i,c} \quad (2)$$

Where, $i$ is the customer; $c$ is the PSC; $\text{Total Contribution Margin Focal Institution}_i$ is the total contribution margin value from the focal institution for all PSCs by customer $i$ at the last month of the training period; $\text{Amount of Credit Focal Institution}_{i,c}$ is the amount of credit taken from the focal institution by customer $i$ per PSC $c$ at the last month of the training period; and $\text{Amount of Credit National Financial System}_{i,c}$ is the amount of credit taken from all financial institutions in the National Financial System by customer $i$ per PSC $c$ at the last month of the training period.

XGBoost Regression was chosen for its superior predictive performance, highlighted by recent studies on gold price forecasting [17] and bankruptcy prediction [18]. Its iterative tree-ensemble approach corrects errors across sequential trees, enhancing accuracy [19]. Bayesian hyperparameter optimization was employed to fine-tune parameters critical to its performance [20]. Model accuracy was evaluated using 10-fold cross-validation, where customer-based subsets were iteratively trained and validated, ensuring robust performance metrics. Predictive accuracy was measured through RMSE and MAE, providing comprehensive insights into model robustness and precision.

### 3.1.2 Phase 2 – Predicting PCM per customer per competitor

In Phase 2, we predict the PCM using the trained model and data shared by competitors through OB protocols. Instead of the focal institution's credit amounts used in training, we utilize credit amounts from competitors' PSCs obtained via OB data (Eq.

3). This approach estimates the PCM the focal institution would generate if all PSCs purchased by customer i from competitor n migrated to it. Multiplying the PCM by 12 projects, the estimated monthly PCM over the next 12 months, assuming the PCM from the last training month, reasonably predicts the upcoming year's value.

$$PCM_{i,n} * 12 \sim Amount\ of\ Credit\ Competitor_{i,c,n} + Amount\ of\ Credit\ National\ Financial\ System_{i,c} \quad (3)$$

Where, $i$ is the customer; $c$ is the PSC; $n$ is the competitor available through OB; $PCM_{i,n}$ is the potential contribution margin for the next 12 months the focal institution would generate if all PSCs purchased by the customer $i$ from competitor $n$ migrated to the focal company at the last month of the training period; $Amount\ of\ Credit\ Competitor_{i,c,n}$ does customer takes the amount of credit $i$ per PSC $c$ from competitor $n$ at the last month of the training period (extracted from OB data); and, $Amount\ of\ Credit\ National\ Financial\ System_{i,c}$: is the amount of credit taken from all financial institutions in the National Financial System by customer $i$ per PSC $c$ at the last month of the training period.

### 3.2 Retention Probability

We used [11] method to address imbalanced retail banking datasets. The framework includes data preprocessing and model training/testing phases. Preprocessing applies feature engineering (FE), ADASYN oversampling, and NEARMISS undersampling. FE generates features like Recency, Frequency, and Monetary Value (RFM) tailored to banking, improving churn prediction. ADASYN creates synthetic churn cases, while NEARMISS removes redundant majority-class instances, balancing the dataset and enhancing the model's ability to detect churn patterns.

In the model training and testing phase, the framework utilizes state-of-the-art classification techniques, specifically the XGBoost model, optimized using Bayesian approaches. The model's performance is evaluated using the precision-recall area under the curve (PR-AUC), accuracy, recall, and specificity. With the prediction of churn, retention can be calculated as Eq. 4.

$$Retention\ Probability_i = 1 - Churn\ Probability_i \qquad (4)$$

Where, $i$ is the customer; $Retention\ Probability_i$ is the probability of retention per $i$; and, $Churn\ Probability_i$ is the Probability of Churn per $i$.

### 3.3 PCLV

The PCLV of the customer $i$ from $n$ competitor is calculated based on Gupta and Lehmann (2005) formula (Eq. 1), assuming a prediction horizon of 12 months, as presented in Eq. 5.

$$PCLV\ Competitor_{i,n} = \frac{(PCM_{i,n} * r_i)}{(1 + d - r_i)} \qquad (5)$$

Where, $i$ is the customer; $n$ is the competitor; $PCM_{i,n}$ is the potential contribution margin for the next 12 months the focal institution would generate if all PSCs purchased by the customer $i$ from competitor $n$ migrated to the focal company at the last month of the training period; $r_i$ is the estimated probability of customer retention of customer $i$ for the next 12 months, and, $d$ is the annual discount rate.

To calculate the PCLV for each customer, we sum the PCLV values for each competitor associated with that customer (eq. 6).

$$PCLV_i = \sum_{n=1}^{N} PCLV\ Competitor_{i,n} \qquad (6)$$

Where, $i$ is the customer; $n$ is the competitor; and, $PCLV\ Competitor_{i,n}$ is estimated in Eq. 5.

### 3.4 Total CLV

Given that the CLV from the focal institution and the PCLV that could migrate from competitors to the focal institution have been estimated for customer $i$, one could eventually calculate the Total CLV of the customer $i$ (Eq. 7).

$$Total\ CLV_i = Traditional\ CLV_i + PCLV_i \qquad (7)$$

Where, $i$ is the customer; $Traditional\ CLV_i$ is the CLV of the customer $i$ from the focal institution; and, $Potential\ CLV_i$ is the CLV of the customer $i$ that could migrate from competitors to the focal institution.

The PCLV estimate depends on OB data availability. As customers consent to share data from more competitors, the focal institution gains a complete view of their total CLV. If a customer shares data from all competitors, the Total CLV will reflect the expected value of their entire wallet.

## 4. DATA AND EMPIRICAL CONTEXT

The proposed approach was tested in a major Brazilian bank with a broad product portfolio limited to credit-related PSC data, representing 73% of the bank's contribution margin. A 36-month dataset (Jan 2021–Dec 2023) covering 261 million transactions for 3,335,034 customers was processed for churn prediction, revealing a significant class imbalance. For PCM prediction, Dec 2023 contribution margins and credit data from the national financial system were used, replacing the bank's credit values with competitor data from OB. However, only 10,507 customers (0.32%) shared OB data, enabling PCM and PCLV predictions exclusively for these customers. A 0.1 annual discount rate was applied, with monetary values converted to USD at R$4.8521/$1 on Dec 31, 2023.

## 5. RESULTS AND DISCUSSION

### 5.1 Model performance

The churn prediction model estimated retention probability as the complement of churn probability, addressing class imbalance during training with 5.4% churn and 94.6% non-churn data. Using 10-fold cross-validation, it achieved a PR-AUC of 0.96, sensitivity of 0.88, and specificity and accuracy of 0.92. Retention probabilities were calculated for all customers, with specific insights for those sharing data via Open Banking (OB).

The XGBoost Regression model's performance in predicting the PCM of the focal institution was also evaluated using 10-fold cross-validation, resulting in a mean RMSE of $74 and MAE of $11.13, demonstrating its predictive accuracy. For context, the contribution margin had a mean of $554.70, a median of $137.12, and a standard deviation of $2,219.00.

## 5.2 PCLV

In Table 2, Actual CLV (Eq. 1) and Aggregated PCLV (Eq. 6) were computed for each customer. Given the PCLV that could be added to each customer's Actual CLV, we summed these two values to reach the Total CLV (Eq. 7). Given this, we compared the segmentation of customers into terciles, following Kumar and Shah (2009), based on Actual CLV and Total CLV. The difference between these two segmentations is that only the PCLV was added by each customer. Given this, we evaluated which customers would migrate among segments in terms of the number of customers and the PCLV added.

Segment analysis revealed three scenarios. The first involves upward transitions, where 4.23% of customers showed potential to move to higher segments (Lower to Intermediate or Upper and Intermediate to Upper), with a $629,183 potential upside (0.95% increase over the Total Actual CLV, which is $66,519,068). The second, 91.56% of customers remained static in the same segment (Upper, Intermediated, and Lower). The PCLV of these customers who remain in the segments indicates a potential upside of $13,145,957 (19.76% increase over the Total Actual CLV). Lastly, we identified 4.20% in downward transitions (Intermediate to Lower and Upper to Intermediate or Lower), valued at $236,953 (0.36% increase over the Total Actual CLV). Given these findings, customers with any upside should be targeted with add-on selling strategies to convert their respective PCLV to the focal institution. Prioritization should rank customers based on their PCLV, focusing on those with higher PCLV.

The accumulated PCLV of all customers resulted in a total of $14,012,093 (an overall upside of 21.06% over the Total Actual CLV). This is significant, especially considering it represents only customers who adopted OB (10,507 customers, which are 0.32% of the 3,335,034 active customers. These results are expected to be much higher as customer adoption of OB increases, showing the relevance of considering the

proposed methods to estimate PCLV to drive marketing efforts at the customer individual level to increase the Actual CLV.

Table 1 – Segmentation change comparing Actual CLV and Potential CLV

| Actual CLV Segment | Total CLV Segment | Total CLV | PCLV | Actual CLV | Number of customers |
|---|---|---|---|---|---|
| Upper | Upper | $67,958,298 (84.36%) | $7,017,311 (50.08%) | $60,940,879 (91.61%) | 3,425 (32.60%) |
| Upper | Intermediate | $195,318 (0.24%) | $40,851 (0.29%) | $154,466 (0.23%) | 75 (0.71%) |
| Upper | Lower | $0 (0%) | $0 (0%) | $0 (0%) | 0 (0%) |
| Intermediate | Upper | $223,812 (0.28%) | $101,601 (0.72%) | $99,781 (0.15%) | 73 (0.70%) |
| Intermediate | Intermediate | $9,136,310 (11.35%) | $4,128,358 (29.46%) | $5,007,952 (7.53%) | 3,061 (29.13%) |
| Intermediate | Lower | $414,087 (0.51%) | $196,102 (1.40%) | $217,985 (0.33%) | 367 (3.49%) |
| Lower | Upper | $0 (0%) | $0 (0%) | $0 (0%) | 0 (0%) |
| Lower | Intermediate | $617,536 (0.77%) | $527,582 (3.78%) | $89,954 (0.14%) | 371 (3.53%) |
| Lower | Lower | $2,008,339 (2.49%) | $2,000,288 (14.27%) | $8,051 (0.01%) | 3,135 (29.83%) |
| **Total** | | **$80,553,700 (100%)** | **$14,012,093 (100%)** | **$66,519,068 (100%)** | **10,507 (100%)** |

## 6. CONCLUSION

Adopting the PCLV framework enables customers' transactions with competitors to be incorporated to assess Total CLV more comprehensively. This allows a focal company to manage Actual CLV, based on internal transaction data, and PCLV, reflecting customers' potential profitability from external transactions. Traditionally, CLV studies have focused solely on the data within the focal company, overlooking the possibility that customers might transact significantly more with competitors. This limited view may not adequately capture the potential profitability of customers, especially those with higher external engagement.

With the advent of Open Banking (OB), financial institutions can now gain insights into customers' external activities, enabling more effective marketing and customer management. Ranking customers by PCLV helps allocate resources to high-impact customers and tailor add-on selling strategies more precisely. OB data also reveals which competitor products customers use, allowing targeted retention efforts for customers with high Actual CLV and low PCLV, ensuring the company maintains its share of their wallet. However, challenges such as low OB adoption and limited competitor transaction data remain. Future research could address these gaps, including developing models to predict customer migration from competitors and enhancing PCLV accuracy.